\documentclass[10pt,letterpaper]{article}
\usepackage{opex3}
\usepackage{cite}
\usepackage{color}
\usepackage{ulem}

\begin{document}
\title{Modification of light transmission channels by inhomogeneous absorption in random media}

\author{Seng Fatt Liew and Hui Cao$^{*}$}

\address{$^1$Department of Applied Physics, Yale University, New Haven, CT 06520, USA\\
}

\email{$^*$hui.cao@yale.edu} 



\begin{abstract}
Optical absorption is omnipresent and very often distributed non-uniformly in space. 
We present a numerical study on the effects of inhomogeneous absorption on transmission eigenchannels of light in highly scattering media. 
In the weak absorption regime, the spatial profile of a transmission channel remains very similar to that without absorption, and the effect of inhomogeneous absorption can be stronger or weaker than homogeneous absorption depending on the spatial overlap of the localized absorbing region with the field intensity maximum of the channel. 
In the strong absorption regime, the high transmission channels redirect the energy flows to circumvent the absorbing regions to minimize loss. 
The attenuation of high transmission channels by inhomogeneous absorption is lower than that by homogeneous absorption, regardless of the location of the absorbing region. 
The statistical distribution of transmission eigenvalues in the former becomes broader than that in the latter, due to a longer tail at high transmission. 
Since the maximum transmission channel is the most efficient in bypassing the absorbing region, the ratio of its transmittance to the average transmittance increases with absorption, eventually exceeds the ratio without absorption. 
\end{abstract}



\section{Introduction}

The ability to control light propagation in turbid media is of great importance to many fields, ranging from medical imaging, laser surgery to photovoltaics and energy-efficient ambient lighting \cite{mosk_NP,sebastien_NC,vos2013,gratzel_JPPC03,hagfeldt_CL10}. 
Thanks to the recent developments of adaptive wavefront shaping and phase recording techniques in optics, the spatial degree of freedom of the input light can be controlled at an unprecedented level of precision. 
These developments have enabled coherent control of light propagation in highly scattering media by manipulating the interference of multiply scattered waves \cite{mosk_NP,vellekoop_PRL, Pendry_view, shi_PRL12, choi_NP, Davy_OE13, sebastien13, vellekoop_OL, changhueiyang_NP, sebastien_PRL,kim_OL13,choi13a}. 
One striking interference effect that has caught much attention is the existence of highly transmitting channels, termed "open channels" in a diffusive system. 
These open channels, which enable an optimally prepared coherent input beam to transmit through a strong scattering medium with order unity efficiency, were predicted initially for electrons \cite{dorokhov1,dorokhov2,Mello88,Mello89, Nazarov94,LamacraftPRB04,AltlandPRL05}. 
Since it is much more difficult to control the input electron states than the input states of classical waves, the wavefront shaping technique has been utilized in the past few years to increase the coupling of the incident light to the open channels of random media \cite{vellekoop_PRL, choi_NP, shi_PRL12, kim_OL13, sebastien13, choi13a}. 
The open channels greatly enhance light penetration into the scattering media, that will have a profound impact in a wide range of applications. 

In reality absorption exists in any material system, and could have a significant impact on light transport in both diffusion regime and localization regime \cite{brouwer, yamilov_OE13, liewPRB14}. 
On one hand the interference effects may be modified by absorption, on the other hand light absorption in strong scattering media can be drastically enhanced or suppressed by interference effects \cite{chong_PRL10, wan_Sci11, chong_PRL11, noh_PRL12, noh_OE13}. 
Thus the interplay between absorption and interference determines not only the amount of energy being transmitted, but also the amount of energy being deposited in a random medium. 
For example, wavefront shaping has enabled focusing of laser light onto a localized absorber that is buried within a random medium to enhance the local absorption \cite{vellekoopOE08}. 

Absorption also has a strong effect on the transmission eigenchannels, especially the open channels. 
The transmission channels are eigenvectors of the matrix $t^{\dagger}t$, where $t$ is the field transmission matrix of the system. 
The eigenvalues $\tau$ are the transmittance of the corresponding eigenchannels. 
In the lossless diffusion regime, the density of the eigenvalues $\tau$ has a bimodal distribution, with one peak at $\tau \simeq 0$ that corresponds to closed channels, and a peak at $\tau \simeq 1$ that corresponds to open channels \cite{dorokhov1,dorokhov2,Mello88,Mello89, Nazarov94,LamacraftPRB04,AltlandPRL05}. 
An open channel has a spatial profile extended throughout the entire random medium, with the intensity maximum near the center \cite{choi_PRB}.
When strong absorption is introduced uniformly across the entire system, the diffusive transport of light in the maximum transmission channel turns into quasi-ballistic \cite{liewPRB14}. 
The straightening of optical paths through the random medium reduces the dwell time and minimize the attenuation by absorption. 
The statistical distribution of transmission eigenvalues are no longer bimodal, as the peak at $\tau \simeq 1$ is diminished by strong absorption \cite{brouwer}. 
Experimentally absorbers are often distributed non-uniformly in random samples, and it is not clear how the open channels would respond to spatially inhomogeneous absorption. 

In this paper, we present a numerical study on the transmission eigenchannels in disordered waveguides with spatially localized absorbing regions. 
We calculate the statistical distribution of transmission eigenvalues and find it can be broader than that with uniform absorption. 
The longer tail of the distribution is attributed to the higher transmission eigenchannels that manage to circumvent the local absorbing regions to minimize loss. 
Compared to other transmission eigenchannels, the one with the largest eigenvalue is the most efficient in bypassing the absorbing regions to transport the maximal amount of energy through the random system. 
Consequently, the ratio of the maximum transmittance to the average transmittance increases with absorption and eventually exceeds the ratio without absorption. 

\section{Numerical model}
In our simulation, we consider a 2D disordered waveguide, shown schematically in Fig. 1(a). 
The dielectric cylinders with refractive index $n$ = 2.0 and radius $r_c$ = 50 nm are randomly positioned inside a waveguide with perfectly reflecting sidewalls. 
The dielectric cylinders occupy an area fraction of 0.04 corresponding to an average distance between cylinders of $a$ = 0.44 $\mu$m. 
The probe light enters the waveguide from the left open end and is scattered by the cylinders. 
The wavelength of input light $\lambda$ is set to 510 nm, to avoid the Mie resonances of individual dielectric cylinders. 
The light is transverse magnetic (TM) polarized, its electric field is parallel to the cylinder axis ($z$-axis). 
The width of the waveguide is $W$ = 10.5 $\mu$m; the number of guided modes in the empty waveguide is $N = 2W/\lambda = 41$. 
The length of the random array of cylinders is $L = $20.6 $\mu$m.  

\begin{figure}[htbp]
	\centering
	\includegraphics[width=\linewidth]{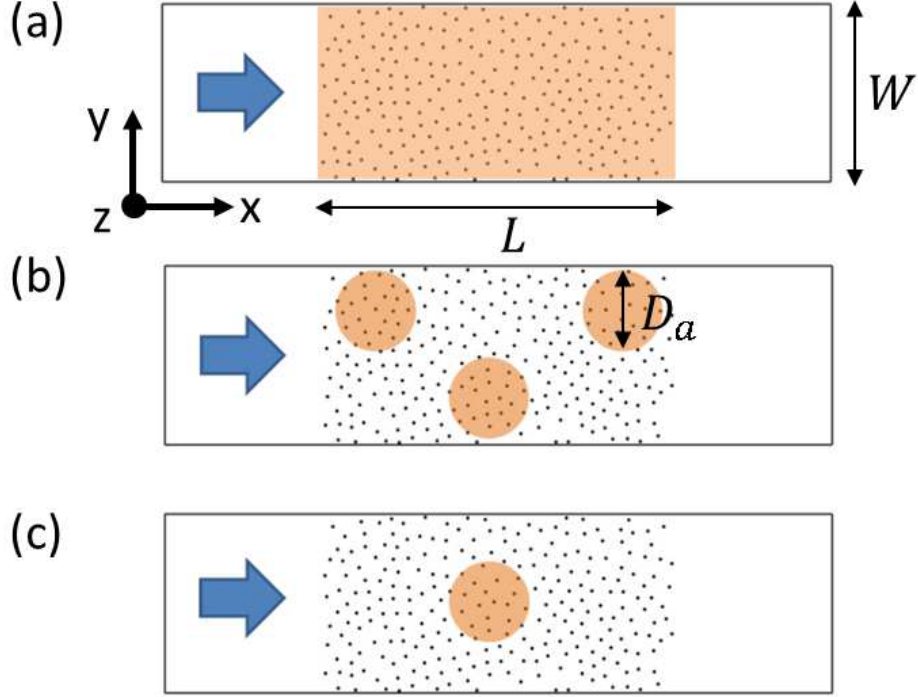}
	\caption{Schematic of the 2D disordered waveguide used in our numerical simulation. Dielectric cylinders are placed randomly in a waveguide with perfect-reflecting sidewalls. (a) $Q_u$: homogeneous distribution of absorbers across the entire random arrays of cylinders. (b) $Q_3$: absorbers are confined to the three circles with diameter $D_a$. (c) $Q_1$: absorbers are concentrated in a single circular region of diameter $D_a$ in the middle of the disordered waveguide. }
	\label{fig1}
\end{figure}

We use the recursive Green's function method \cite{leePRL81,beenakkerPRL96, chong_PRL11} to calculate the transmission matrix of the disordered waveguide \cite{liewPRB14}, which gives the output field for any arbitrary input. 
Using the input and output fields as the boundary conditions, we further compute the field distribution inside the disordered waveguide. 
The field intensity is averaged over the waveguide cross-section to give the evolution $I(x)$ along the waveguide (in the $x$ direction). 
The ensemble-averaged $\langle I(x) \rangle$ displays a linear decay, from which we extract the transport mean free path $l_t =$ 1.65 $\mu$m \cite{rossumRMP99}. 
The localization length is then estimated to be $\xi = (\pi/2)Nl_t = $107 $\mu$m. 
Since $l_t \ll L \ll \xi$, the propagation of light in the disordered waveguide can be described by diffusion. 

After characterizing the scattering properties, we introduce optical absorption to the disordered waveguide.  
Below we consider three cases. 
The first one, labeled $Q_u$, has homogeneous absorption across the entire random structure [Fig. \ref{fig1}(a)]. 
More specifically, a constant imaginary refractive index, $n_i>0$, is introduced to both dielectric cylinders and background, to avoid any additional scattering caused by the spatial inhomogeneity of $n_i$. 
The ballistic absorption length is $l_a = 1/(2kn_i) = 1/(\rho\sigma_a)$, where $k = 2\pi/\lambda$ is the wavevector, $\rho$ is the density of absorbers, and $\sigma_a$ is the absorption cross-section of each absorber. 
The diffusive absorption length, given by $\xi_a = \sqrt{l_tl_a/2}$, determines the strength of absorption effects.
In the weak absorption regime, $\xi_a >L$, the average length of diffusive paths inside the random medium  $l_p = 2L^2/l_t$ is shorter than $l_a$, thus most scattering paths are barely affected by absorption.
In the strong absorption regime, $\xi_a < L$, large attenuation of long scattering paths significantly modifies the transport through the system \cite{liewPRB14}. 

The next two cases have non-uniform absorption in the disordered waveguides, one is labeled $Q_3$ where the absorption is confined to three isolated circles [Fig. \ref{fig1}(b)], the other is $Q_1$ which has a single absorbing region in the middle of the waveguide [Fig. \ref{fig1} (c)]. 
All the circular absorbing regions have the same diameter $D_a = 2.7l_t$,  and the space in between the absorbing regions, as well as the distance from an absorbing region to the waveguide sidewall, is larger than the transport mean free path. 
When comparing the effects of absorption in the above three cases, the total number of absorbers is kept constant, so that only the spatial distribution of absorbers is different.
Since $n_i$ is proportional to the density of absorbers $\rho$, the product $n_i \, S_a$ is the same, where $S_a$ is the total area of absorbing regions. 
The smaller the $S_a$, the larger is the value of $n_i$, and the stronger is the absorption within the absorbing region. 
Hence, $l_a \propto S_a$ and $\xi_a \propto \sqrt{S_a}$.  
Below, the absorption strength is given by $\alpha L/\xi_a$, where $\alpha = \sqrt{S_a/(LW)}$ is a scaling factor which is equal to $0.27$ for $Q_1$, $0.46$ for $Q_3$ and $1.0$ for $Q_u$. 

\section{Statistical distribution of transmission eigenvalues}

\begin{figure}[htbp]
	\centering
	\includegraphics[width=\linewidth]{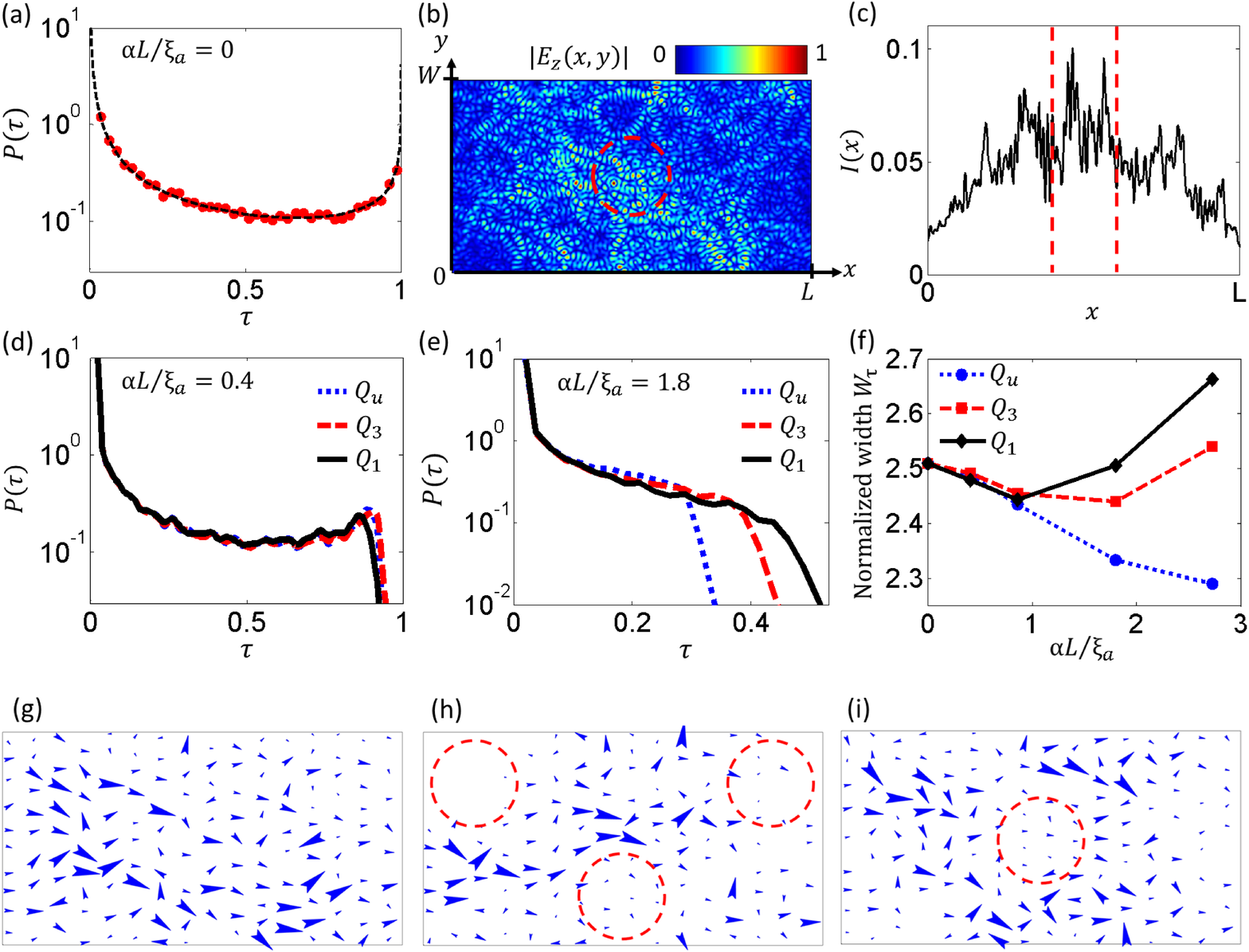}
	\caption{Statistical distribution of transmission eigenvalues $P(\tau)$ for disordered waveguide with homogeneous and inhomogeneous absorption. 
	(a) Without absorption, $P(\tau)$ (filled circles) exhibits the bimodal distribution given by Eq. (2) (dashed line). 
	(b) Spatial distribution of the electric field amplitude $|E_z(x,y)|$ of the highest transmission eigenchannel in one random realization of disordered waveguide. 
	(c) Cross-section-averaged intensity along the $x$-direction, having the maximum at the center of the waveguide, which coincides with the absorbing region in $Q_1$ (marked by the vertical dashed lines). 
	(d) $P(\tau)$ for weak absorption $\alpha L/\xi_a = 0.4$ [$\alpha = 1$ for $Q_u$, $\alpha = 0.46$ for $Q_3$ and $\alpha = 0.27$ for $Q_1$]. The peak near $\tau=1$ is diminished and shifted towards smaller $\tau$. The shift in the case of inhomogeneous absorption is slightly larger than that of homogeneous absorption, as the open channels experience more attenuation due to better spatial overlap with the localized absorbing regions. 
	(e) $P(\tau)$ for strong absorption $\alpha L/\xi_a = 1.8$. $Q_1$ has the highest transmission among the three cases. 
	(f) The normalized width $W_{\tau}$ of $P(\tau)$ as a function of the absorption strength $\alpha L/\xi_a$. 
	(g, h, i) Spatial map of the normalized Poynting vector $\vec{S}'(x,y)$ for the maximum transmission channel in the disordered waveguide with homogeneous absorption $Q_u$ (g), three absorbing regions $Q_3$ (h) and one absorbing region $Q_1$ (i) at $\alpha L/\xi_a = 1.8$. Dashed circles in (h, i) mark the boundary of the absorbing regions. When absorption is strong and inhomogeneous, main energy flows bypass the absorbing regions.}
	\label{fig2}
\end{figure}

A singular value decomposition of the transmission matrix $t$ gives
\begin{eqnarray}
t = U\  \Sigma \ V^{\dagger} \ , 
\end{eqnarray} 
where $\Sigma$ is a diagonal matrix with non-negative real numbers,  $\sigma_n = \sqrt{\tau_n}$, $\tau_n$ is the transmittance of the $n^{th}$ transmission eigenchannel, $\tau_1 > \tau_2 > \tau_3 ... > \tau_N$. 
$U$ and $V$ are $N \times N$ unitary matrix, $V$ maps input channels of the empty waveguide to eigenchannels of the disordered waveguide, and $U$ maps eigenchannels to output channels. 
The input singular vector that corresponds to the highest transmission eigenvalue $\tau_1$ gives the maximum transmission eigenchannel, its elements represent the complex coefficients of the waveguide modes that combine to achieve the highest transmission through the random waveguide.

In the absence of absorption, the density of the transmission eigenvalues of the disordered waveguide $P(\tau)$ [Fig. \ref{fig2}(a)] follows the  bimodal distribution [dashed line in Fig. \ref{fig2}(a)]: \cite{dorokhov1,dorokhov2,Mello88,Mello89,Nazarov94,LamacraftPRB04,AltlandPRL05}
\begin{eqnarray}
P(\tau) = \frac{\bar{\tau}}{2}\frac{1}{\tau \sqrt{1-\tau}}, 
\end{eqnarray}
where $\bar{\tau}$ represents the average transmittance. 
As shown in Fig. 2(a), the distribution has two peaks, one at $\tau \simeq 1$ and another at $\tau \simeq 0$. 
The transmission eigenchannels with $\tau \simeq 1$ are ``open channels", and the ones at $\tau \simeq 0 $ are ``closed channels". 
The diffusive transport is dominated by the open channels, and $\bar{\tau}$ is determined by the number of open channels \cite{dorokhov1,dorokhov2}. 
Figure \ref{fig2}(b) plots the spatial distribution of field amplitude $|E_z(x,y)|$ for the maximum transmission channel in one random realization of the disordered waveguide. 
The input light penetrates through the entire waveguide. 
The cross-section-averaged intensity $I(x) = (1/W) \int_0^W |E_z(x,y)|^2 dy$, shown in Fig. \ref{fig2}(c), is peaked at the center of the waveguide \cite{choi_PRB}. 

When absorption is introduced uniformly across the disordered waveguide, the open channels experience more attenuation than the closed channels since light in an open channel propagates deeper into the waveguide. 
Consequently, the peak of $P(\tau)$ near $\tau=1$ is diminished and shifted towards smaller $\tau$ [dotted line in Fig. \ref{fig2}(d)].
The effect of absorption can be further enhanced by inhomogeneous absorption, e.g. in the case of $Q_1$, all absorbers are concentrated in the center of the waveguide, which coincide with the intensity maximum of the highest transmission channel [dashed line in Fig. 2(c)]. 
The calculated $P(\tau)$, shown in Fig. 2(d), has the $\tau\simeq 1$ peak moved farther to smaller $\tau$, reflecting a faster decreasing transmission of the open channels in $Q_1$. 
 
However, as we continue increasing absorption, the behavior changes completely. 
As seen in Fig. \ref{fig2}(e), the peak of $P(\tau)$ at large $\tau$ vanishes in all three cases, and the bimodal distribution is replaced by a monotonic decay of $P(\tau)$ with $\tau$.   
Surprisingly, the disordered waveguide with a single absorbing region $Q_1$ exhibits the longest tail at high transmission, followed by the waveguide with three separate absorbing regions $Q_3$, while the waveguide with uniform absorption $Q_u$ has the shortest tail. 
This trend is just opposite to that with weak absorption. 
To quantify the change in the width of $P(\tau)$, we plot the normalized width $W_{\tau} = \sqrt{\langle \tau^2\rangle/\langle \tau\rangle^2 -1}$ in Fig. \ref{fig2}(f). 
While $W_{\tau}$ for $Q_u$ decreases monotonically with increasing absorption, $W_{\tau}$ for $Q_3$ and $Q_1$ first decreases and then increases, eventually exceeds the value without absorption. 

\section{Maximum transmission channel}

To understand why inhomogeneous absorption can reach higher transmission than homogeneous absorption, we examine the maximum transmission channel in the presence of strong absorption. 
To map the energy flow inside the disordered waveguide, we compute the Poynting vector $\vec{S}(x,y) = (1/2) \textnormal{Re}[\vec{E}(x,y)\times \vec{H}^*(x,y)$]. 
The net flow over a cross-section of the disordered waveguide is $F(x) = \int_0^W S_x(x,y) dy$, where $S_x(x,y)$ is the projection of  $\vec{S}(x,y)$ on the $x$-axis.
While the net flux $F(x)$ remains constant in the absence absorption, it decays in the presence of absorption. 
For a clear visualization of the energy flow deep inside the random structure, the Poynting vector $\vec{S}(x,y)$ is normalized by $F(x)$,  $\vec{S}'(x,y) = \vec{S}(x,y)/F(x)$.
Figure \ref{fig2}(g-i) plot the normalized Poynting vector $\vec{S}'(x,y)$ for the maximum transmission eigenchannels in the disordered waveguide with homogeneous or inhomogeneous absorption. 
With homogeneous absorption $Q_u$ [Fig. \ref{fig2}(g)], light propagates mostly in the forward direction in order to reduce the dwell time within the random waveguide to minimize loss. 
Light transport changes from diffusive to quasi-ballistic when absorption is strong \cite{liewPRB14}. 
In disordered waveguide with three separate absorbing regions $Q_3$ [Fig. \ref{fig2}(h)], the main energy flows are in between the three separate absorbing regions to avoid the absorption. 
When there is only one absorbing region in the middle of the waveguide $Q_1$ [Figure \ref{fig2}(i)], the incoming energy stream splits into two to circumvent the absorbing region. 
Therefore, the high transmission channels have modified their energy flows to bypass the absorbing regions in the waveguide so as to achieve higher transmission in the case of inhomogeneous absorption than that of homogeneous absorption. 

\begin{figure}[htbp]
	\centering
	\includegraphics[width=\linewidth]{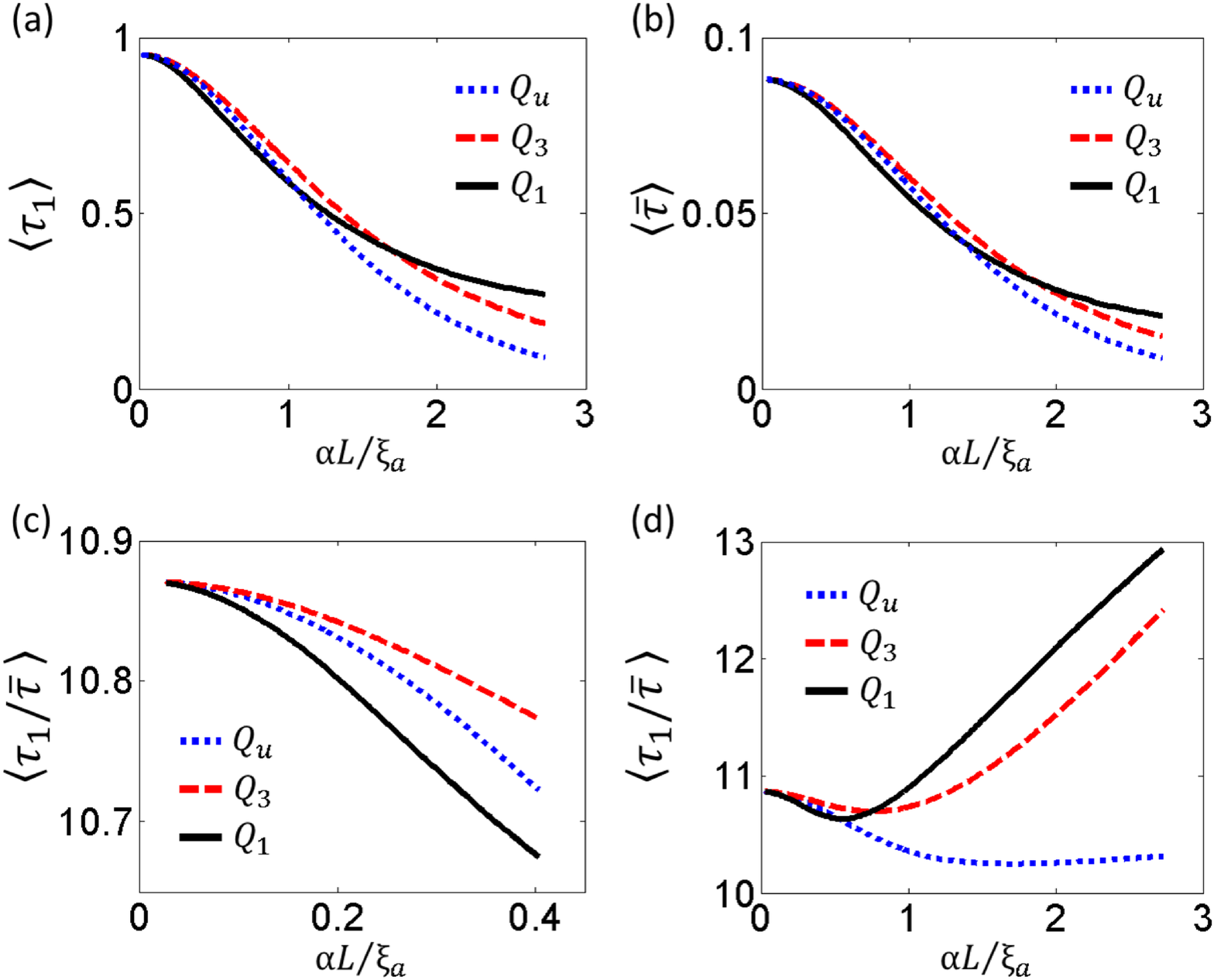}
	\caption{Comparison of the maximum transmission eigenvalue $\tau_1$ and the average of all transmission eigenvalues $\bar{\tau}$ in the presence of inhomogeneous absorption to that of homogeneous absorption. 
	(a) $\tau_1$ decreases with absorption in all three cases. The drop is the fastest for $Q_1$ when absorption is weak [$\alpha L/\xi_a < 1$] but switches to the slowest when absorption is strong [$\alpha L/\xi_a > 1$]. 
	(b) Similar trends are observed for the change of $\bar{\tau}$ with absorption. 
	(c) The ensemble-averaged ratio $\langle \tau_1/\bar{\tau} \rangle$ shows the fastest reduction for $Q_1$ compared to $Q_u$ and $Q_3$ in the weak absorption regime. 
	(d) $\langle \tau_1/\bar{\tau} \rangle$ starts to increase in the strong absorption regime, and eventually exceeds the ratio without absorption in $Q_1$ and $Q_3$. }
	\label{fig3}
\end{figure}

Next, we track the evolution of the highest transmission channel with a gradual increase of absorption. 
Figure \ref{fig3}(a) plots the change in the maximum transmission eigenvalue $\tau_1$ as a function of the absorption strength $\alpha L/\xi_a$ for the disordered waveguide with homogeneous or inhomogeneous absorption. 
When absorption is weak $\alpha L/\xi_a < 1$, $\tau_1$ for $Q_1$ decreases faster than $Q_3$ and $Q_u$. 
However, as absorption becomes strong $\alpha L/\xi_a > 1$, the drop of $\tau_1$ slows down and its value for $Q_1$ is higher than the other two cases. 
This transition occurs when the maximum transmission channel modifies its energy flow to bypass the absorbing region. 
For comparison, we plot the average of transmission eigenvalues $\bar{\tau}$ as a function of absorption strength in Fig. \ref{fig3}(b). 
Its decay is qualitatively similar to that of ${\tau}_1$, because $\bar{\tau}$ has the largest contribution from $\tau_1$. 
Their ratio $\tau_1/\bar{\tau}$ displays the subtle difference in the change of their decay rates by absorption. 
In the weak absorption regime [Fig. \ref{fig3}(c)], the ensemble-averaged ratio $\langle \tau_1/\bar{\tau} \rangle$ decreases monotonically for all three cases. 
The reduction is the fastest for $Q_1$ when the number of total absorbers is the same. 
This can be understood from the results in the previous section. 
The spatial distribution of the maximum transmission channel is barely modified by weak absorption, and it experiences the most attenuation when all absorbers are concentrated in the spatial location where its field intensity is maximal ($Q_1$). 
The lower transmission channels do not penetrate as deep into the random waveguide, thus their intensity maxima are closer to the input end of the wavewguide. 
The less spatial overlap with the absorbing region leads to lower attenuation, thus the average of transmission eigenvalues $\bar{\tau}$ decreases more slowly than $\tau_1$. 

With a further increase of absorption, the ratio $\tau_1/\bar{\tau}$ starts increasing, especially in the case of inhomogeneous absorption ($Q_3$, $Q_1$), its value eventually exceed that without absorption. 
This is because the maximum transmission channel bypasses the absorbing regions to reduce the loss, and its transmittance decreases less than other eigenchannels. 
Thus the reduction of $\tau_1$ becomes smaller than that of $\bar{\tau}$, leading to an increase of their ratio $\tau_1/\bar{\tau}$ with absorption. 
The ratio $\tau_1/\bar{\tau}$ surpasses its value without absorption at $D_a/ \xi_a^{(3)} \approx 0.6$ for $Q_3$ and $D_a/ \xi_a^{(1)} \approx 0.8$ for $Q_1$, where $\xi_a^{(3)}$ and $\xi_a^{(1)}$ are the diffusive absorption lengths within the absorbing regions in $Q_3$ and $Q_1$ respectively. 
Hence, the highest transmission eigenchannel starts to circumvent the absorbing regions when the diffusive absorption length becomes comparable to the size of the absorbing region. 
In the case of homogeneous absorption, $\tau_1/\bar{\tau}$ levels off and then increases slightly when $\alpha L/\xi_a > 1$ [Fig. \ref{fig3}(d)]. 
Since the absorbers are everywhere in the random waveguide, light in the maximum transmission channel cannot bypass absorbers when transmitting through the waveguide. 
The only way to minimize loss is to shorten the dwell time inside the waveguide by taking less winding paths when absorption is strong. 
The other transmission channels do not change as much as the maximum transmission channel, so they may experience slightly higher loss. 
 
\section{Other transmission eigenchannels}

\begin{figure}[htbp]
	\centering 
	\includegraphics[width=\linewidth]{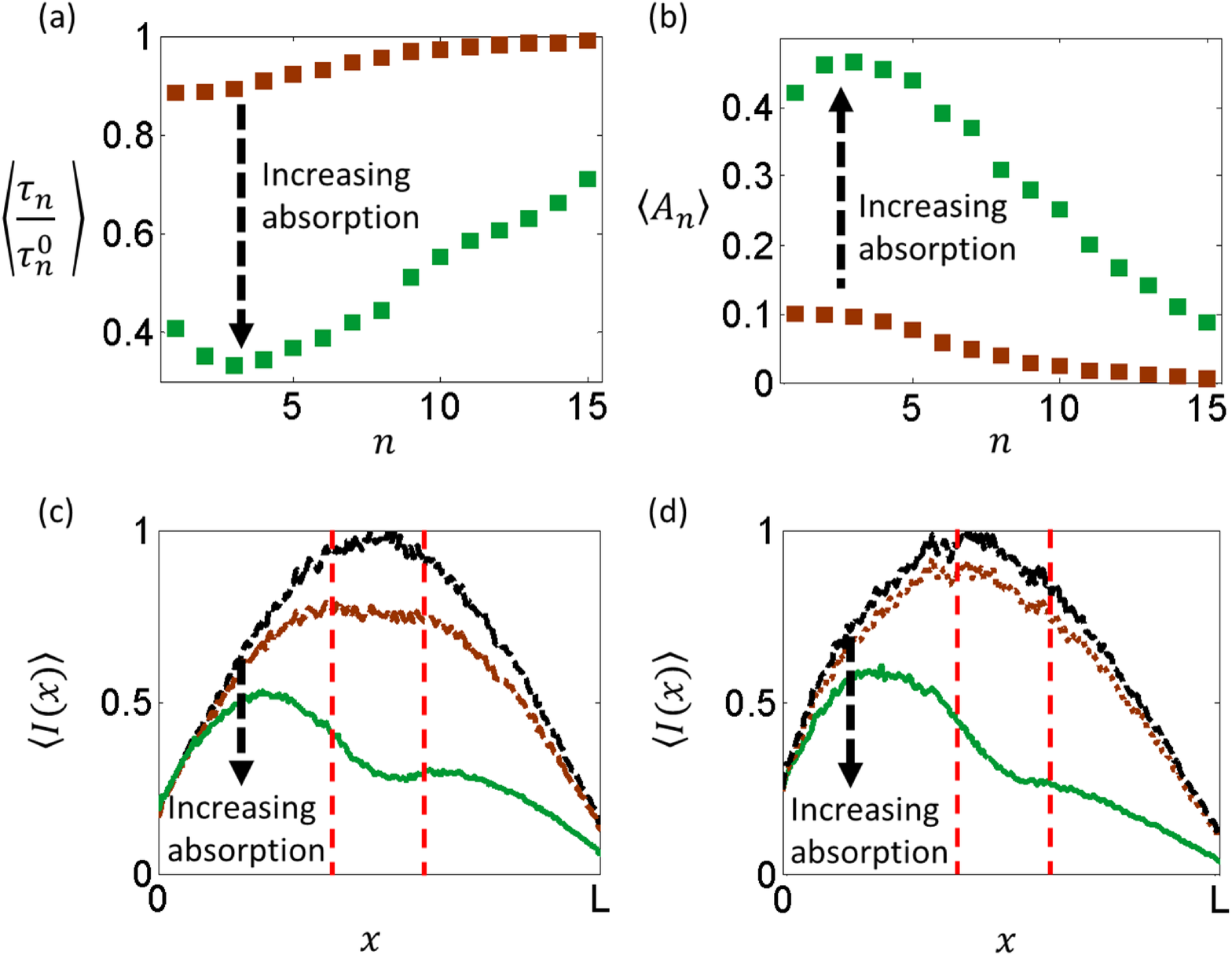}
	\caption{Modification of transmission eigenchannels in disordered waveguide with a single absorbing region in the middle $Q_1$. (a) Ensemble-averaged ratio $\langle \tau_n/ \tau_n^0 \rangle$ for individual eigenchannels. The reduction of $\tau_1$ is the largest when absorption is weak  $D_a/\xi_a^{(1)} = 0.32$ but becomes less than other eigenchannels such as $\tau_3$ when absorption is strong $D_a/\xi_a^{(1)} = 1.45$. 
	(b) The absorption of individual eigenchannels $A_n$. The first eigenchannel is the one that experiences the most absorption at weak absorption but it is replaced by the third eigenchannel at higher absorption level. 
	(c, d) Ensemble average of cross-section-averaged electric field intensity $\langle I(x)\rangle$ for the first and third eigenchannels. Both are normalized to their maximal values when there is no absorption. With weak absorption $D_a/\xi_a^{(1)} = 0.32$ (brown dotted lines), the field intensity of the first eigenchannel decreases more at the waveguide center than the third eigenchannel. The vertical dashed lines mark the boundary of the absorbing region. When absorption is strong $D_a/\xi_a^{(1)} = 1.45$, the reduction of field intensity behind the absorbing region becomes less for the first eigenchannel than the third eigenchannel as the former is modified to circumvent the absorbing region. }
	\label{fig4} 
\end{figure}

In this section, we investigate the changes in other transmission eigenchannels due to inhomogeneous absorption. 
Figure \ref{fig4}(a) shows the ratio $\tau_n/\tau_n^0$ for $Q_1$, where $\tau_n$ is the transmission eigenvalue for the $n^{th}$ channel with absorption and $\tau_n^0$ is the eigenvalue without absorption. 
When absorption is weak $D_a/\xi_a^{(1)} <1$, $\tau_1$ decreases faster than all others. 
We also calculate the absorption in each channel, $A_n = 1-\tau_n-R_n$, where $R_n$ is the reflectance of the $n^{th}$ transmission eigenchannel. 
The absorption in the maximum transmission channel $A_1$ is the largest  [Fig. \ref{fig4}(b)], which explains why $\tau_1$ reduces more than other channels when absorption is weak. 
However, at higher absorption level $D_a/\xi_a^{(1)} >1$, the transmission reduction of other eigenchannels such as $\tau_3$ exceeds that of $\tau_1$. 
Correspondingly, the absorption experienced by the third eigenchannel $A_3$ becomes larger than the first eigenchannel $A_1$ [Fig. \ref{fig4}(b)]. 

The above result can be explained by the modification of these transmission channels by inhomogeneous absorption. 
In a lossless disordered waveguide, the field intensity maximum of the first transmission eigenchannel is located at the center of the waveguide, whereas the intensity maximum of the third eigenchannel is shifted towards to the input end of the waveguide [Fig. \ref{fig4}(c, d)]. 
When absorption is introduced only to the central region of the waveguide, the field intensity of the first channel is peaked in the middle of the absorbing region while the field intensity of the third channel is peaked at the edge [black dashed lines in Fig. \ref{fig4}(c, d)]. 
With weak absorption $\xi_a^{(1)} > D_a$, the spatial distribution of field intensity in both channels are barely changed, thus the first eigenchannel is attenuated more than the third eigenchannel [brown dotted lines in \ref{fig4}(c, d)]. 
However, at strong absorption where the diffusive absorption length within the absorbing region becomes shorter than the size of the absorbing region $\xi_a^{(1)} < D_a$, light transmission through the central absorbing region becomes very low and the scattering paths in the first eigenchannel avoid that region to achieve high transmission as seen in Fig. \ref{fig2}(i). 
The circumvention of light paths around the absorbing region helps to reduce the absorption of light and therefore the field intensity in the section of the waveguide behind the absorbing region is higher for the first eigenchannel than the third eigenchannel [green solid lines in Fig. \ref{fig4}(c, d)]. 
 
Similar modifications of the transmission eigenchannels are observed in the disordered waveguide with three separate absorbing regions $Q_3$. 
Since it is more difficult to bypass three absorbing regions than a single one, the decrease of $\tau_1$ is higher for $Q_3$ than for $Q_1$ [Fig. \ref{fig3}(a)], and also $Q_3$ has a narrower distribution of transmission eigenvalues than $Q_1$ [Fig. \ref{fig2}(e,f)]. 
Note that in both $Q_1$ and $Q_3$, the space in between the absorbing regions or between the absorbing region and the waveguide wall is larger than the transport mean free path, so that the multiple scattering of light in the non-absorbing regions and the interference of the scattered light enable an efficient steering of energy flow away from the absorbing regions. 

\section{Conclusion}

We have performed a detailed numerical study to understand how spatially non-uniform absorption modifies the transmission eigenchannels in a 2D disordered waveguide. 
In the weak absorption regime, the spatial profile of a transmission channel remains very similar to that without absorption, and the effect of inhomogeneous absorption can be stronger or weaker than homogeneous absorption depending on the spatial overlap of the localized absorbing region with the field intensity maximum of the channel. 
In the strong absorption regime, the high transmission channels redirect the energy flows to circumvent the absorbing region to minimize loss. 
Thus the attenuation of high transmission channels by inhomogeneous absorption is lower than that by homogeneous absorption, making the statistical distribution of transmission eigenvalues in the former broader than that in the latter. 
Compared to other transmission eigenchannels, the one with the largest eigenvalue is the most efficient in bypassing the absorbing regions to transport the maximal amount of energy through the random medium. 
Hence, the ratio of the maximum transmittance to the average transmittance, as well as the normalized width of the eigenvalue distribution, increases with absorption and eventually exceeds the value without absorption. 
Our numerical study provides a physical understanding of the effects of inhomogeneous absorption on transmission eigenchannels in diffusive media. 
The finding that inhomogeneous absorption may have a weaker impact on open channels than homogeneous absorption is promising for practical applications. 

\section{Acknowledgement}
We thank Allard Mosk, Willem Vos, Alexey Yamilov, Arthur Goetschy, S\'{e}bastien Popoff, and Patrick Sebbah for stimulating discussions. 
The numerical study is supported in part by the facilities and staff of the Yale University Faculty of Arts and Sciences High Performance Computing Center. 
This work is funded by the US National Science Foundation under the Grant No. DMR-1205307 and ECCS-1068642.

\end{document}